\documentclass[11pt]{article}
\usepackage{moriond,epsfig}
\usepackage{amssymb}
\usepackage{hyperref}

\def\bea{\begin{eqnarray}}
\def\eea{\end{eqnarray}}

\begin{document}

\title{$D$-decays with unquenched Lattice QCD}

\author{ Benjamin Haas }

\address{Laboratoire de Physique Th\'eorique (Bat. 210) \\ Universit\'e Paris--Sud, Centre d'Orsay \\ F--91405 Orsay-Cedex, France \\ E-mail: {\tt Benjamin.Haas@th.u-psud.fr} }

\maketitle
\abstracts{
We discuss the recent progress in computing the $D$--meson decay constant  and $D\rightarrow \pi \ell \nu$ form factors from the lattice 
QCD simulations with $N_f=2$ dynamical  using Wilson quarks.
We report $f_D=201(20)~{\rm MeV}$ and $F_+(1~{\rm GeV}^2)/f_D=4.04(78)~{\rm GeV}^{-1}$ at $a\simeq0.08~{\rm fm}$.
}

\section{Introduction and lattice setup}
Accurate determinations of the Cabibbo--Kobayashi--Maskawa (CKM) couplings provide an essential test of  the Standard Model. 
The most straightforward method for their extraction is through the leptonic and/or semileptonic meson decays. However, such an extraction
 from the experimently measured decay widths requires a reliable information about the hadronic quantities, 
namely the decay constants and the form factors. Such an  information  is expected to be provided by the QCD simulations on the lattice.

Many lattice calculations of these quantities have been attempted over the past twenty years. Until a few of years ago, all 
computations were performed  in the {\it quenched} approximation in which  the effect of virtual quark loops is neglected.
  Recent progress 
allowed us to move to the  {\it unquenched}  case in which at least $N_f=2$ dynamical quarks are present in the QCD vacuum fluctuations.
 The lattice  quark action that are being used 
nowadays are ${\cal O}(a)$--improved so that the systematic errors are  ${\cal O}\left((am_c)^2\right)$.
 In this write-up we present new results for the charmed decays using  an improved Wilson  action
with  $N_f=2$ degenerate sea quarks. The results reported here refer to the  simulations made at  $a\simeq 0.08~{\rm fm}$ for 
three different values of the  sea quark mass corresponding to $m_{\pi_{qq}}\simeq 770~{\rm MeV},~600~{\rm MeV}~{\rm and}~400~{\rm MeV}$.~\cite{Gockeler:2005rv}
Other unquenched results relevant to the leptonic and semileptonic decays were obtained by using the staggered quark action~\cite{Aubin:2004ej}. Since
the formal proof of validity  of the staggered formulation is still missing, the study based on Wilson quarks is more than needed.

\section{Leptonic decays}
\label{sec-lpt}
\subsection{Hadronic matrix element}
The simplest way to determine the CKM matrix element $\left|V_{cd} \right|$ is via the leptonic decay $D\rightarrow \ell \nu$ with $\ell=\tau,\mu,e$.    The decay width  is given by
\bea
\label{eq-dw-lept}
\Gamma(D^+\rightarrow \ell^+ \nu_\ell)   =\left|V_{cd}\right|^2  \frac{G_F^2}{8\pi} m_{D^+}m_\ell^2 \left(1-\frac{m_\ell^2}{m_{D^+}^2} \right){f_{D}}^2,
\eea
where  $G_F$ is the  Fermi  constant and the decay constant $f_D$  parametrizes the hadronic matrix element:
\bea\left \langle 0 |A_\mu \vert D(p) \right \rangle = i f_D p_\mu \;,
\eea
with $A^\mu=\overline{c}\gamma_\mu \gamma_5 d$. The theoretical uncertainty in eq. (\ref{eq-dw-lept}) is entirely due to $f_D$. For $\ell=e$ or $\mu$ this decay mode has been recently accurately measured~\cite{Artuso:2005ym}.
On the lattice, $f_D$ is extracted from the asymptotic  behavior of the  $2$--point Green function, i.e.
\bea
C^{(2)}_\mu (t)=\sum_{\vec{x}} \left \langle 0| \left(A_\mu\right)_{\vec{0},0} \left(\bar{c}\gamma_5q\right)_{\vec{x},t} |0 \right \rangle
\stackrel{t\gg 0}{\longrightarrow} \left \langle 0| A_\mu\vert D(\vec{0}) \right \rangle\times \frac{\sqrt{{\cal Z}_D}}{2m_D} e^{-m_Dt}\; ,
\eea
where $A_\mu$ is the  appropriately renormalized axial current.  ${\cal Z}_D$ is evaluated  from:
\bea
\sum_{\vec{x}} \left \langle 0| \left(\bar{c} \gamma_5q\right)_{\vec{0},0} \left(\bar{c}\gamma_5q\right)_{\vec{x},t} |0 \right \rangle
\stackrel{t\gg 0}{\longrightarrow} \frac{{\cal Z}_D}{2m_D} e^{-m_Dt}\; .
\eea

\subsection{Computation of  $f_D$ and chiral extrapolations}
\begin{figure}
\begin{center}
\includegraphics[width=7cm]{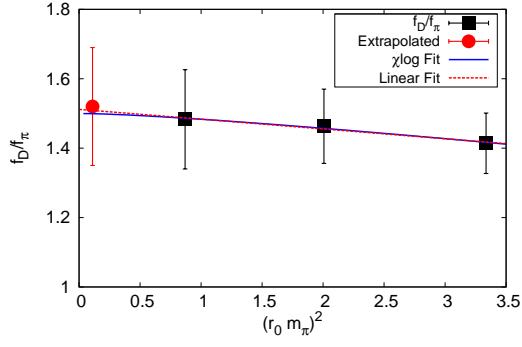}
\end{center}
\caption{The extrapolation of the ratio $f_D/f_\pi$ either by using the HM$\chi$PT formula, eq. (\ref{eq-fdfp}) (solid line) or a linear fit (dashed line). \label{fig-fD} }
\vspace{-0.6cm}
\end{figure}
We focus to the unquenched case in which the valence and the sea quark masses are equal. Thus, for each of our directly 
accessible light quark masses, we compute $f_{D_q}$ which then needs to be extrapolated to the physical $f_{D_d}\equiv f_D$. That extrapolation can be made either linearly in $m_q$, or by using the
 expression derived in  heavy meson chiral perturbation  theory (HM$\chi$PT)~\cite{Casalbuoni:1996pg}:
\bea
\label{eq-fdcpt}
f_{D_q}\sqrt{m_{D_q}} &=& \Phi_0\left[1-\frac{3}{4}{\frac{1+3g^2}{\left(4\pi f_0\right)^2}m_{\pi_{qq}}^2~{\rm log}~m_{\pi_{qq}}^2}+c_\Phi m_{\pi_{qq}}^2 \right].
\eea
In this formula,  $\Phi_0$ and $c_\Phi$  are the fit parameters,  $f_0$ is the  pion decay constant in the chiral limit   determined 
on the same lattice, while $g$  is related to the coupling  between the
heavy meson doublet, $(D,D^\star)$ and the soft pion, i.e.
\bea
\left\langle D(k)\pi(q)|D^\star(p,\lambda) \right\rangle=(\epsilon^\lambda.q)g_{D^\star D \pi}= (\epsilon^\lambda.q)\frac{2\sqrt{m_Dm_D^\star}}{f_\pi}g\;,
\eea
were $\epsilon^\lambda$ is the polarization vector of $D^\star$. Its value has been previously computed in the static limit~\cite{Abada:2003un} ($g\simeq0.5$), and with  the propagating charm quark~\cite{Abada:2002xe} ($g\simeq 0.6$), both values leading
to the large  factor  multiplying the logarithm in eq. (\ref{eq-fdcpt}),  $3/4(1+3g^2)$. That would drive the extrapolation
of $f_{D_q}\sqrt{m_{D_q}}$ way below the result obtained through the linear extrapolation, thus to  large systematic errors.  It is then
  more convenient to consider the  ratio $f_D/f_\pi$ in which  the chiral logarithmic term is halved:   \pagebreak[4]
\bea
  \frac{f_{D_{q}}}{f_{\pi_{qq}}} &=& \frac{\Phi_0}{f_0\sqrt{m_{D_q}}} \left[1+\frac{1}{4}{\frac{5-9g^2}{\left(4\pi f_0\right)^2}m_{\pi_{qq}}^2~{\rm log}~m_{\pi_{qq}}^2}+c_1 m_{\pi_{qq}}^2 \right] \;,
\label{eq-fdfp}
\eea
$c_1$ being a fit parameter. From fig. (\ref{fig-fD}), we see that linear and $\chi$--log fits are very consistent and by using both  
values for $g$ mentioned above, we get $\left.f_D/f_\pi=1.50(24)\right._{{\chi}{\rm-log}}$ and  
$\left. f_D/f_\pi=1.52(17)\right._{{\rm linear}}$. With the physical pion decay constant $f_{\pi}=139.6~{\rm MeV}$ \cite{Yao:2006px}, we arrive at
\bea
f_D=201(22)\left(^{+4}_{-9} \right)~{\rm MeV} \;.
\eea
\vspace{-1.3cm}
\section{Semileptonic decays}
\label{sec-slpt}
\subsection{Hadronic matrix element}
\begin{figure}
\includegraphics[width=8cm]{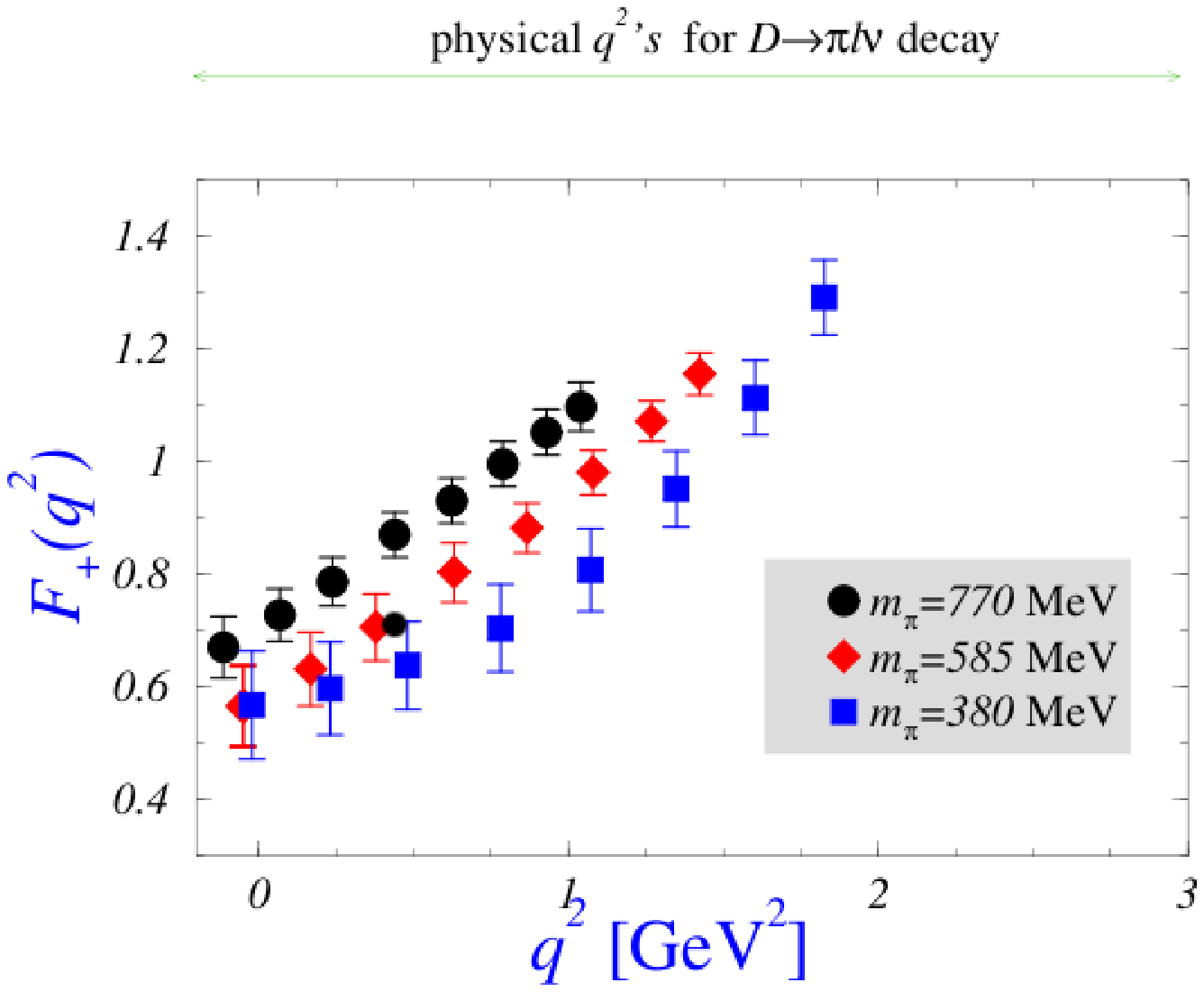}
\includegraphics[width=8cm]{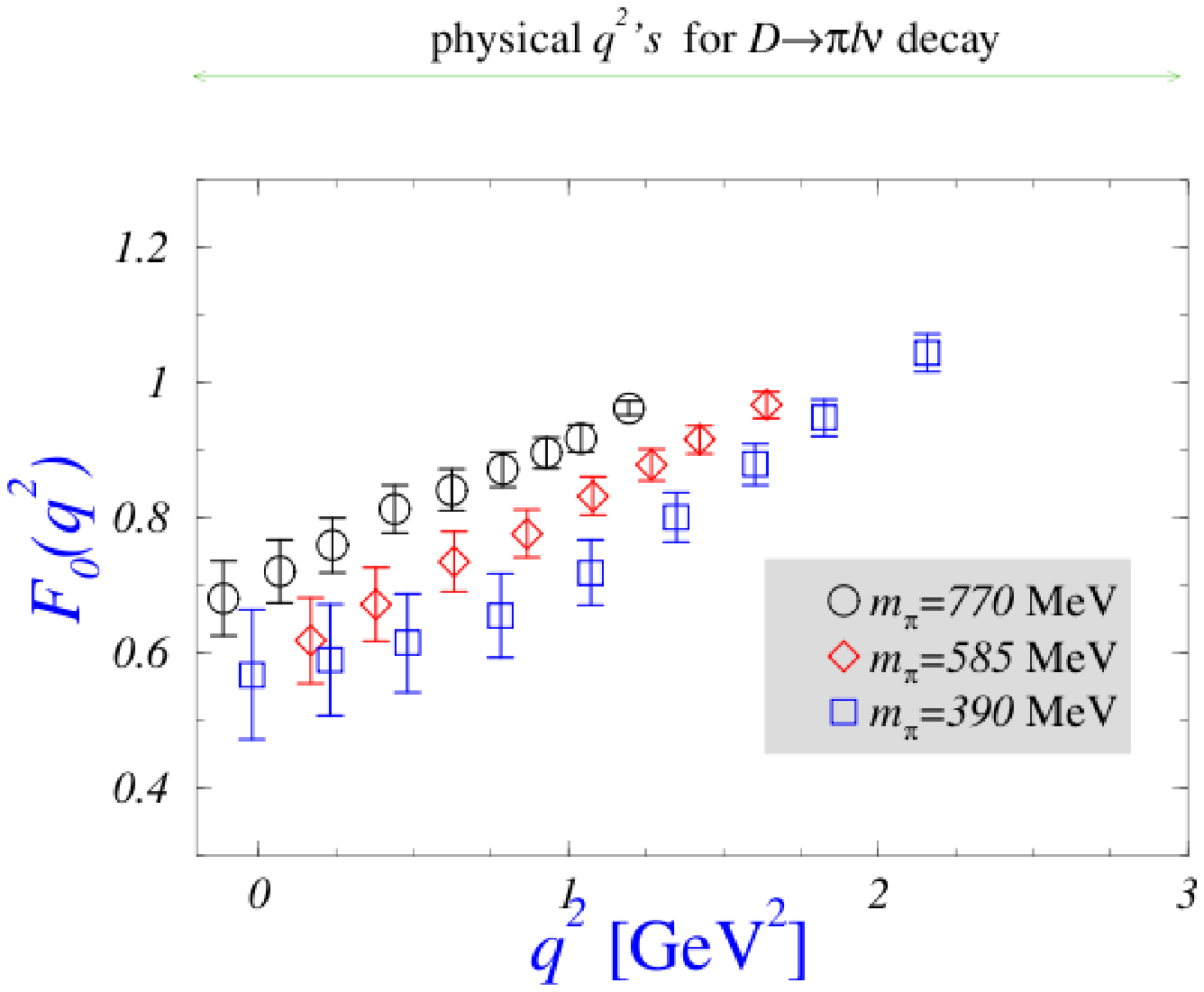}
\caption{The $q^2$-dependence of the vector (left) and the scalar (right) form factors relevant to $D\rightarrow \pi \ell \nu$ decay for 3 different pion masses, accessible directly from our lattices.  \label{fig-FF}}
\end{figure}

 $\left|V_{cd}\right|$ can also be extracted by studying the partial or total decay width of the semileptonic decay $D\rightarrow\pi\ell\nu_\ell$ ($\ell=e,\mu$), which has been measured in various recent experiments  \cite{EXP-Semilept}.  The differential decay width is given by 
\bea
\label{equ-dslp}
\frac{d\Gamma}{dq^2}(D\rightarrow \pi \ell \nu_\ell)   =|V_{cd}|^2 \frac{G_F^2}{192 \pi^2m_{D}^3}\lambda^{3/2}\left(q^2\right) \left|F_+(q^2)\right|^2\; ,
\eea
where $q$ is the momentum transfer and  $\lambda(q^2)=(q^2+m_D^2-m_\pi^2)^2-4m_D^2m_\pi^2$. The  {\it vector}  form factor $F_+\left(q^2\right)$ parametrizes the hadronic matrix element of the weak current
\bea
\left \langle \pi (\vec{k}) \vert \left(V-A\right)_\mu \vert D(\vec{p}) \right \rangle=\left(p+k-q\frac{{m_D}^2-{m_\pi}^2}{q^2} \right)_\mu F_+\left(q^2\right)+q_\mu\frac{{m_D}^2-{m_\pi}^2}{q^2}  F_0\left(q^2\right)\;,
\eea
where $q=p-k$ and $F_+(0)=F_0(0)$.  The  contribution of the {\it scalar} form factor  $F_0\left(q^2\right)$ to the decay width comes with a  $m_\ell^2$--factor and therefore can be neglected. 
The form factors are extracted from the  behavior of the following $3$--point correlation functions:
\bea \nonumber
C^{(3)}_{\mu}(\vec{k},\vec{q},t;t_s)&=\sum_{\vec{x},\vec{y}} \left\langle 0|\left(\overline{q}\gamma_5q \right)_{\vec{0},0}  \left(V_\mu \right)_{\vec{y},t} \left(\overline{c}\gamma_5q \right)_{\vec{x},t_s} e^{-i\left( \vec{k}\vec{y}-\vec{q}\vec{x} \right)}|0\right\rangle\\& \stackrel{0\ll t \ll t_s}{\longrightarrow} \frac{\sqrt{{\cal Z}_\pi}}{2E_\pi} e^{-E_\pi t}\left \langle \pi(\vec{k}) \vert V_\mu \vert D(\vec{p}) \right \rangle \frac{\sqrt{{\cal Z}_D}}{2E_D} e^{-E_D (t_s - t)} \; ,
\eea
computed on the lattice.  The axial current  does not contributed due to the parity conservation in QCD. Extracting the matrix element from a fit of $C^{(3)}_{\mu}$ requires the knowledge of several independent  quantities (${\cal Z}_{\pi,D}$, renormalization constants and ${\cal O}(a)$ improvement coefficients of the vector current) whose error   lower the accuracy of $F_+\left(q^2\right)$. This can be improved by using the strategies  we previously studied in \cite{Becirevic:2007cr}, based on ratios of such $3$--point correlators and twisted boundary conditions. In fig. \ref{fig-FF} we show the results obtained with the $1^{\rm st}$ strategy. 
\subsection{Computation of $F_+\left(q^2=1~{\rm GeV}^2\right)$ and chiral extrapolations}
The differential decay width in eq. (\ref{equ-dslp}) is experimentally measured for various values of $q^2$. To extract 
$\vert V_{cd}\vert$, one then needs  a single $F_+\left(q^2\right)$--value.  We choose  $q^2=1~{\rm GeV}^2$ where both theoretical  and experimental errors  are under a
reasonable control.  The extrapolation of $F_+(1~{\rm GeV}^2)$  toward the physical value leads to a
 difficulty similar to what we discussed above.   To get around that difficulty, we use HM$\chi$PT 
fits for the ratio  $F_+\left(q^2\right)/f_D$  where large deviations due to $m_\pi^2 \log m_\pi^2$ terms are 
reduced~\cite{Becirevic:2003ad}. We get

\bea \nonumber
\frac{F_+(1~{\rm GeV}^2)}{f_D} = 4.32(56)~{\rm GeV}^{-1} \; {\rm (HM\chi PT~fit)} \quad {\rm ;} \quad3.76(54)~{\rm GeV}^{-1} \;\ {\rm (Linear~fit)} \; ,
\eea
where we also quote our result obtained by using the naive linear extrapolation. The difference of the two is an
 estimate of the systematic uncertainty of the extrapolation procedure. 
\section{Summary}
In this short note, we reported  on the progress in determining the key hadronic quantities entering the  leptonic and semileptonic 
decays on the lattice by using the Wilson quarks. 
A better control over the systematic uncertainties is achieved  if one  chooses judiciously the ratios in which various sources of uncertainties  cancel out, or are diminished. 
In the case of semileptonic decays, also the use of twisted boundary conditions is very important.  The quenched
experience suggests that the ${\cal O}(a^2)$ artifacts are reasonable~\cite{Juttner:2003ns}, but to that end, more work is needed.
\section*{Acknowledgments}
We thank the QCDSF collaboration for letting us use their gauge field configurations and 
the \textit{Centre de Calcul de l'IN2P3 \`a Lyon}, for giving us access to their computing facilities.

\end{document}